\input epsf
\documentstyle[preprint,aps]{revtex}
\large
\begin{document}
\draft

\title
{Channel Interference in a Quasi Ballistic Aharonov-Bohm Experiment}

\author{G. Cernicchiaro$^{1,4}$, T. Martin $^2$, K. Hasselbach$^1$, D.
Mailly$^3$, A. Benoit$^1$}
\address{$^1$ CRTBT-CNRS, 25 av des Martyrs, 38042 Grenoble, France}
\address{$^2$ CPT-Universit\'e d'Aix-Marseille II, 163 av de Luminy, 13288
Marseille, France}
\address{$^3$ LMM-CNRS, 196 av. H. Ravera, 92220 Bagneux, France}
\address{$^4$CBPF-CNPq, 150 av. Xavier Sigaud, 22000 Rio de Janeiro, Brazil}
\maketitle
\begin{abstract}
New experiments are presented on the transmission of electron waves through
 a 2DEG (2 dimensional electron gas) ring with a gate on top of one of the branches. 
Magnetoconductance 
oscillations are observed, and the phase of the Aharanov-Bohm signal alternates 
between $0$ and $\pi$ as the gate voltage is scanned. A Fourier transform of the 
data reveals a dominant period in the voltage which corresponds to the 
energy spacing between successive transverse modes.
A theoretical model including random phase shifts between successive modes reproduces 
the essential features of the experiment.
\end{abstract}
\smallskip
\bigskip
\narrowtext

The Aharonov--Bohm (AB) effect
has proven to be an invaluable tool
for quantifying interference phenomena in mesoscopic physics.
Early experiments on long metal cylinders \cite{Sharvin}
revealed that an electron accumulates a phase $\int {\bf A}.d{\bf l}$
as it is scattered elastically by impurities 
while traveling around the loop: when the magnetic flux is varied,
an oscillatory pattern with periodicity $h/2e$ results from
the interference of an electron wave with its time reversed path
\cite{Altshuler}.
Experiments on gold loops \cite{Webb} confirmed that for normal metals
which are laterally confined, the expected periodicity
\cite{BILP} is that
of the {\it single} flux quantum $\phi_0=h/e$. The amplitude
of the magnetoresistance background can be understood
within the framework of universal conductance fluctuations (UCF)
\cite{UCF}.
The two dimensional electron gas (2DEG) formed at the heterojunction between two 
semiconductors is used for experiments in the ballistic transport regime.
Recently, two experiments on gated rings in the diffusive 
\cite{Webb gate}, and the ballistic \cite{Yacoby91}
regime reported oscillations associated with the modulation 
of the electron wave length under the gate. 

In the present letter, results on a new AB
transport measurement in the ballistic regime are exposed.
The number of lateral channels in one branch of the ring is
adjusted by means of an electrostatic gate.
In addition to the usual AB interference, a periodic pattern
is uncovered when the number of lateral electron modes is modulated
with the gate. 
It is emphasized that the conductance pattern contains two
types of oscillations: a) the wavelength variation under the gate
\cite{Webb gate,Yacoby91}, and b) a ``new'' periodicity associated
with the closing/opening of transverse modes in the ring.
Phase switching and period halving in the AB pattern is monitored as the
confinement is varied. The essential features of the data can be
interpreted using  the scattering formulation of quantum
transport \cite{BILP}. The inclusion of
disorder is necessary to explain the alternation of AB phases.

The 2D electron gas was created at the interface of
a GaAlAs/GaAs heterojunction. Via standard electron beam
lithography \cite{Mailly}
a single loop device  of width $1.2 \mu m$ with inner
diameter $4 \mu m$, connected to measurement
leads was designed (Fig. 1). In this etched structure, the width of the
wire which constitutes the ring is further reduced by a
depletion on each edge of $0.27 \mu m$. The 2DEG presents
 a mobility of $1.14\times 10^{6}cm^{2}$/Vs, an electron density of
$n_s=3.6\times 10^{11}cm^{-2}$. The coherence
length $l_\phi>20 \mu m$ and the mean free path
$l_e = 11,3 \mu m$, are consistent with the ballistic regime.
A metallic gate was deposited over one branch of the ring (hereby referred
to as the upper branch) allowing a controlled
depletion of the 2DEG underneath.
The number of electron channels $N$ in the wires
defining the ring is estimated
assuming parabolic 
confinement in the transverse
direction. The width $W$ of the channel roughly equals
the ratio of the 1D to the 2D electron density \cite{Berggren}.
For $W=600 nm$, and a Fermi wave length
$\lambda_F=40 nm$,  $N=(3\pi/4) W/\lambda_F \approx 30$ channels.
The lateral dimensions of the wire are
comparable to that of the conductance quantization experiments
\cite{vanWees}, where the number of 
transmitted channels was shown to scale linearly with 
the depletion voltage. 

The conductance is determined using standard synchronous detection
measurements in a four terminal configuration. A low-frequency AC current
of 10 nA was injected and measurements were taken at 15 mK, in a dilution
refrigerator. An external flux variation of 12 gauss was applied
corresponding to 4 flux quanta in the mean radius of the ring.
While lowering the gate voltage from 0 to - 300 mV by 1 mV steps, the
complete conductance pattern was measured over a period of 4 hours.
Digital filtering routines were applied to reduce base line variations due to UCFs.
In Fig. 2,  a ``landscape plot'' of the data is displayed: The periodicity of the 
AB signal survives
until a voltage of about $-250 mV$ where the electrons underneath 
the gate are completely 
depleted and the ring is effectively cut off.
When both arms transmit electrons, 
shaded and clear areas alternate
in the vertical direction, indicating
oscillatory behavior as a function the gate voltage.
Attention is focused on the alternating contrast and the phase reversal when the gate 
voltage is increased. The smoothly changing background is identified as a residue of the 
total UCF signal.
The phase of the AB pattern takes only the values
$0$ or $\pi$ \cite{Yacoby95}. The absence of asymmetry 
under field reversal \cite{Benoit} is attributed to the large distance (further than 
$l_\phi$) between the pair of 
current and voltage terminals on each side of the ring.

In the inset of Fig. 3, a magnetoresistance trace is displayed ($V_g=0V$),
 and the corresponding Fourier signal shows a dominant
component at the single flux quantum $h/e$.
Higher harmonics have a much reduced amplitude. The signal
corresponding to total depletion ($V_g=-300mV$) in the upper branch
is plotted for comparison (square symbols). 
In Fig. 4, the modulus of the $h/e$ Fourier component
of the ring {\it resistance} is plotted for each value of $V_g$.
The location of the transition between maxima and 
minima matches the position of the contrast changes in 
the pattern of the conductance landscape (Fig. 2).
While the location of the resistance peaks appears to be 
chaotic, a detailed analysis (see below) reveals 
a regular structure. Peaks of reduced magnitude persist between
$-200 mV$ and the depletion voltage.

The Fourier transform of 
the $h/e$ harmonic is computed for several intervals (width $128 mV$), starting from 
different initial voltages (Fig. 5).
A dominant peak at $0.062 (mV)^{-1}$
(arrow, center of figure)
corresponding to a voltage period of
$16 mV$ appears in all intervals. 
Smaller peaks beyond $0.13 (mV)^{-1}$ correspond to  
higher harmonics. 
Below $ 0.062(mV)^{-1}$, additional
peaks (arrow, left of Fig. 5)
are observed, which shift in position from one 
voltage interval to the next, from $ 0.037 (mV)^{-1}$ (Fig. 5a) to $ 0.015 (mV)^{-1}$
(Fig. 5d). 
This shift reflects the changing barrier 
height as the mean gate voltage increases \cite{Webb gate,Yacoby91}: 
Oscillations, originating from several channels, correspond
to an integer number of Fermi wave lengths
$(2m(E_n-V))^{-1/2}$ 
($E_n$ subband energy, $V$ barrier height)
over the length of the gate. 

On the other hand,
the robustness of the $16 mV $ oscillation in all intervals signifies that 
it has a different origin. Lateral confinement
gives rise to conductance quantization 
in the upper branch of the ring, and the resulting magnetoresistance 
signal reflects how many channels in the upper branch, 
contribute to the interference pattern. Between $0$ and
$-250 mV$, we detect approximately $16$ periodic peaks in Fig. 4.
Clearly this number is lower than the estimated number of channels.
Nevertheless the presence of a metallic gate induces changes in the electrical potential 
underneath, even at zero applied voltage (a positive voltage of a few hundred millivolts is 
necessary to open all channels \cite{Mailly}).
While it is not possible to pinpoint 
exactly how many channels contribute to each oscillation 
of Fig. 4, the number of oscillations is
comparable to the estimated number of transmitting channels.

Theoretically, the scattering approach
for coherent transport \cite{BILP} 
predicts a conductance:
\begin{equation}
G=2{e^2\over h}\sum_n |s_n|^2~,
\label{landauer}
\end{equation}
where the sum is taken over the transverse electron modes,
and the $|s_n|^2$ are the eigenvalues of the {\it transmission} matrix
multiplied by its hermitian conjugate. 
The splitting of the waves between
the upper and lower branch (Fig. 1), is prescribed 
for each mode by a $3\times 3$ scattering matrix 
\cite{Buttiker Azbel}. No reflection is assumed in 
the lower branch. A scatterer with a quantized 
conductance \cite{Buttiker quantization} is located 
on the upper branch.
Phases $\theta_n$ are added to the waves crossing 
the gate: disorder corresponds to random phases.
Electron waves are then recombined into the collecting lead, 
and the accumulated phase differences lead to an interference pattern.
For a minimal description, no mixing between channels 
(backscattering) is 
introduced explicitly. Nevertheless, the symmetry
of the S--matrix implicitly allows scattering
between incoming (outgoing) channels on each side.
The transmission coefficient for electrons waves
in the $n$--th channel is given by:
\begin{eqnarray}
s_n&=&-2\epsilon(\sqrt{1-2\epsilon}+1)^{-2}e^{i\pi\phi/\phi_0}
\nonumber\\
& &\times
\left[\begin{array}{cc}
1& 1\end{array}\right]
\left[{\bf t}_l{\bf t}_g(n){\bf t}_le^{2i\pi\phi/\phi_0}-{\bf 1}\right]^{-1}
\left[\begin{array}{c}
1 \\ -1
\end{array}\right],
\label{transmission coefficient}
\end{eqnarray}
where $\epsilon$ controls the connection of the
current probes to the ring ($\epsilon_{min}=0$ no coupling to the ring;
$\epsilon_{max}=1/2$ for optimal coupling). ${\bf t}_l$ (${\bf t}_g$) is a
$2\times 2$ {\it transfer} matrix
describing the beam splitters on both sides of the ring 
(the gate which controls transmission in the upper branch). 
Calculations of the conductance landscape, ($6$ channels), are plotted in
Fig. 6. In Fig. 6a ($\theta_n=0$ for all channels) the 
conductance pattern displays a 
staircase structure which 
results from the progressive opening/closing of the 
channels in the upper branch.
A succession of parallel ``crests'' and ``valleys'',
illustrates that the AB signals of all channels are in phase. 
A periodic variation of the phase 
shifts, $\theta_{n+1}-\theta_n=\pi$ \cite{Geraldo} leads 
to a halving of the AB period \cite{Yacoby96} for specific voltages. 
Peaks of reduced size arise at these locations. The phase of the AB signal at zero flux is 
unperturbed.

Finally, disorder is introduced by choosing random shifts.
Several configurations of disorder were tested, with 
the following conclusions:
For a substantial depletion, peaks with a high conductance 
appear despite the reduction of the number of 
transmitting channels.
The phase of the AB signal alternates between
the two values $0$ and $\pi$ when the voltage is swept. 
In Fig. 6b, phase shifts $[0,\pi,\pi,0,0,\pi/2]$ were 
picked so as to highlight the pertinent features:
The landscape 
contains an alternation of peaks shifted by $\pi$
with respect to their nearest neighbors. By subtracting
a linear background to Fig. 6b, a periodicity is observed
in {\it both} the magnetic flux and the depletion.
Finally, peak to valley variations of the conductance 
in Fig. 6b occur on a relatively small voltage scale 
(roughly 2 conductance steps of Fig. 6a).

In conclusion, channel interference and conductance quantization are shown to explain the 
essential features of this Aharonov Bohm experiment.
When both arms are transmitting, the suppression of 
{\it one} single channel triggers fluctuations of the 
magnetoresistance which are comparable to the average signal,
and may operate a shift of $\pi$ of the pattern.
Strikingly, the regular structure in the 
pattern persists from large depletion voltages, 
where few modes propagate in the lower branch
(the typical conditions of Ref. \cite{vanWees}), to
zero depletion voltages where most channels are
transmitted.
Our calculations reveal that random phase disorder must be included in the model to 
obtain the main features of the experiment, such as the sudden
horizontal shifts of the pattern and the large variations of the AB signal  
when the number of channels is reduced.
Physically, this phase disorder 
originates from variations in path lengths 
suffered by different modes, from geometric scattering 
at the beam splitters, or from 
inhomogeneities in the confinement (impurities, etc....). 

This work sheds further light on the issue of sudden 
phase changes in interference experiments. 
We speculate that this geometry can be used 
for the quantitative study of other scatterers
located on a branch of the ring.
The analysis of the resulting AB pattern then 
provides a way to quantify the role of disorder
and geometric scattering. Is the pattern
altered significantly when the sample is 
heated (new impurity configuration), or it is 
tied primarily to the confinement ? 
In particular, it may prove interesting 
to study rings in which a quantum billiard is 
embedded.
Finally, the 
addition of a gate to modulate the channels
is also useful for the magnetic response of 
{\it isolated} rings. It may allow to scan through 
different ring configurations and therefore
to vary the persistent current from a paramagnetic
to a diamagnetic signal.
\indent
G.C. acknowledges support by the CNPq, Brazil.

\pagebreak
\medskip
\begin{figure}
\caption{Atomic force microscope image: detail of a sample with the gate (white 
regions) over the
upper branch and the gates for the leads.}
\end{figure}

\begin{figure}
\caption{AB component of the conductance, measured as a
function of applied flux $\phi$ (horizontal axis) and gate voltage (vertical axis). Areas of 
dark (clear) contrast indicate minima 
(maxima) in the AB signal.}
\end{figure}

\begin{figure}
\caption
{Inset: Flux dependence of the sample resistance, for a closed 
($V_g= 0V$) and open ($V_g=-300mV$) ring. Main figure: FFT transforms of these
signals after subtraction of an offset.}
\end{figure}

\begin{figure}
\caption{ Absolute value of the h/e Fourier component of the A-B resistance as function 
of gate voltage.}
\end{figure}

\begin{figure}
\caption{ The Fourier transforms of $R_{h/e}$ (FIG. 4.) over an $128mV$ interval for 
different initial voltages ($a\to d$) as indicated. The persistent peak corresponds to a 
voltage periode of $16mV$.}
\end{figure}

\begin{figure}
\caption{ a) Calculated total transmission trough an Aharonov-Bohm ring with $6$ modes, 
$\theta_n=0$, as a function of 
flux and gate voltage. b) Same calculation with channel shifts $[0,\pi,\pi,0,0,\pi/2]$.
}
\end{figure}

\end{document}